\documentclass[aps,twocolumn,showpacs,pre,10pt]{revtex4-2}
\usepackage{epsfig,amsmath,color,colordvi,verbatim}
\usepackage[normalem]{ulem}
\usepackage{braket}
\usepackage{float}
\usepackage{hyperref}

\usepackage{romannum} 
\AtBeginDocument{\pagenumbering{arabic}}

\def\be{\begin{equation}}
\def\ee{\end{equation}}
\def\bee{\begin{eqnarray}}
\def\eee{\end{eqnarray}}

\begin{document}
\title{Unraveling Geometric-phase at Conical Intersection by Cavity-enhanced Two-dimensional Electronic Spectroscopy} 

\author{Yang-Cheng Ye$^{1}$}
\author{Fulu Zheng$^{1}$} \email{zhengfulu@nbu.edu.cn} 
\author{Ajay Jha$^{2,3,4}$}\email{Ajay.Jha@rfi.ac.uk} 
\author{Hong-Guang Duan$^{1}$}\email{duanhongguang@nbu.edu.cn}
\affiliation{
$^1$Department of Physics, School of Physical Science and Technology, Ningbo University, Ningbo, 315211, P.R. China \\
$^2$Rosalind Franklin Institute, Harwell, Oxfordshire OX11 0QX, United Kingdom\\
$^3$Department of Pharmacology, University of Oxford, Oxford, OX1 3QT United Kingdom \\
$^4$Research Complex at Harwell, Rutherford Appleton Laboratory, Didcot OX11 0QX, United Kingdom 
}
\date{\today}

\begin{abstract}
The geometric phase is a fundamental quantum mechanical phenomenon uniquely associated with conical intersections (CIs) between potential energy surfaces and serves as a definitive signature of their presence. In this study, we propose a novel spectroscopic approach to directly detect the geometric phase (GP) using two-dimensional electronic spectroscopy (2DES) enhanced by strong light-matter interactions within an optical cavity. Focusing on a prototypical pentacene dimer undergoing singlet fission, we model the nonadiabatic wave packet dynamics as it evolves through a CI. The optical cavity enables dynamic modulation of the coupling between the optical field and molecular vibrational modes, allowing precise control over the wave packet pathways. Importantly we identify a cancellation in the spectral amplitude arising from phase differences accumulated along different trajectories, which serves as a clear spectroscopic manifestation of the GP. A pathway-resolved analysis shows that the GP-induced wave packet interference is directly reflected in cavity-dependent excited-state absorption cross-peaks in 2DES. This cavity-enhanced 2DES framework not only enables direct observation of GP effects but also offers a versatile platform for probing ultrafast nonadiabatic processes. Our results provide fundamental insights into topological effects in molecular dynamics and pave the way for experimental strategies in quantum control, photochemistry, and the design of advanced optoelectronic materials.

\end{abstract}

\maketitle

\maketitle


A conical intersection (CI) is a point where two potential energy surfaces (PESs) become degenerate, typically induced by the coupling of two nuclear coordinates \cite{Domcke book1, Domcke book3, Mukamle Chem Rev}. This degeneracy results in strong nonadiabatic coupling in the vicinity of the CI, which causes the Born-Oppenheimer approximation to break down \cite{BO approx}. These CIs act as funnels, enabling ultrafast non-radiative transitions between electronic states, typically on timescales of tens to hundreds of femtoseconds \cite{NatPhys CI}. CIs play a central role in key molecular processes such as photoisomerization \cite{Massimo ChemSocRev}, internal conversion \cite{PNAS Nucleobases, ChemSci CI}, and singlet fission \cite{Kukura Nature Phys 2015, Duan Sci Adv 2020}, which are critical to vision, photosynthesis, and emerging photonic technologies. Understanding their dynamics is essential for accurate modeling of molecular behavior. As such, CIs are a focus of theoretical and computational studies aiming to inform the design of advanced photoactive systems. 

Initially deemed theoretical curiosities due to the extreme conditions required for their observation, CIs were long considered elusive in real molecules. Advances in ultrafast spectroscopic methods have since provided compelling evidence of their existence and significance in diverse polyatomic systems. Mathies and colleagues employed stimulated Raman spectroscopy to track rhodopsin photoisomerization, showing sub-200 fs electronic deactivation \cite{Science Mathies 2005}, while Tahara and colleagues demonstrated ultrafast structural evolution in cis-stilbene, implicating CI-mediated dynamics \cite{Science Tahara 2008}. Subsequent transient grating studies by Miller and colleagues revealed even faster isomerization ($<$ 50 fs) \cite{Nature Chem Miller 2015}, and two-dimensional electronic spectroscopy (2DES) experiments further unveiled intricate excited-state couplings \cite{JPCB Philip 2017}. Moreover, evidence increasingly links CIs to singlet fission processes \cite{Kukura Nature Phys 2015, Duan Sci Adv 2020, Musser Nature Comm 2019, Matsumoto Nature Chem 2017}. Notably, light-induced CIs have recently emerged as a focus of study \cite{JPCL 12 2052 2021, PCCP 23 16868 2021, AVS Quant Sci 6 023501 2024, JPCL 9 1951 2018}. Understanding their mechanistic nature is therefore vital for elucidating  the nonadiabatic dynamics that govern ultrafast photochemical reactivity and energy relaxation. 

Theoretical investigations have shown that nonadiabatic dynamics near CIs can be effectively modeled using a simplified two-state-two-mode framework \cite{Domcke1, Olivucci PNAS 2000, Domcke3}, where the molecular structural changes are represented by tuning and coupling coordinates \cite{JCP 151 244102 2019, JCP 151 014106 2019}. Maxim {\em et al.} explored these dynamics by simulating time-resolved 2DES, demonstrating that the evolution of wave packets traversing the CI can be discerned through excited-state absorption (ESA) features \cite{Maxim JPB 2014}. Other theoretical studies further support the utility of 2DES \cite{2DES1, 2DES2} as a powerful technique to probe the complex dynamics surrounding CIs \cite{Duan JPCL1, Duan JPCL2, Egorova CP}. Extending beyond the optical regime, recent efforts have also proposed exploring these nonadiabatic phenomena using ultrafast X-ray spectroscopy \cite{Mukamel1, Mukamel2, Mukamel3, Kowalewski1}. However, it has become evident that rapid electronic deactivation alone does not unambiguously confirm the presence of a CI, because similar ultrafast behavior can arise in systems with avoided crossings (ACs) \cite{Mukamel4}. In light of this, Duan {\em et al.} compared CI and AC models, revealing that only CIs exhibit a geometric phase (GP) -- a topological feature arising from the wave function's phase change upon encircling the intersection \cite{Duan CP 2018}. This distinction positions the GP as a definitive signature of a CI, providing a crucial criterion for experimentally distinguishing CIs from other nonadiabatic phenomena. Although this phase offers a theoretically rigorous fingerprint for identifying CIs, its experimental detection remains elusive. The subtlety and complexity of observing the GP in real molecular systems highlight a critical gap in current methodologies. Thus, while the GP holds promise as a definitive signature of CIs, realizing its practical detection continues to pose a formidable challenge, underscoring the need for more refined experimental strategies and theoretical tools aimed at isolating and measuring this topological hallmark.

This study presents a theoretical framework for directly detecting the GP associated with CIs via 2DES enhanced by an optical cavity. By modulating the light-matter coupling strength, we isolate spectroscopic features uniquely arising from GP-driven wave packet dynamics. A four-level pentacene dimer model incorporating two vibrational modes and a CI between excited electronic states enables clear observation of these effects on a symmetric potential energy surface. We identify magnitude cancellation in ESA as a distinct spectroscopic signature of the GP. The sensitivity of ESA to cavity coupling offers a tunable detection mechanism. This work establishes a versatile framework for probing quantum dynamics near CIs and provides a pathway toward controlled exploration of molecular topological phenomena and coherence.


\begin{figure}[t!]
\begin{center}
\includegraphics[width=8.5cm]{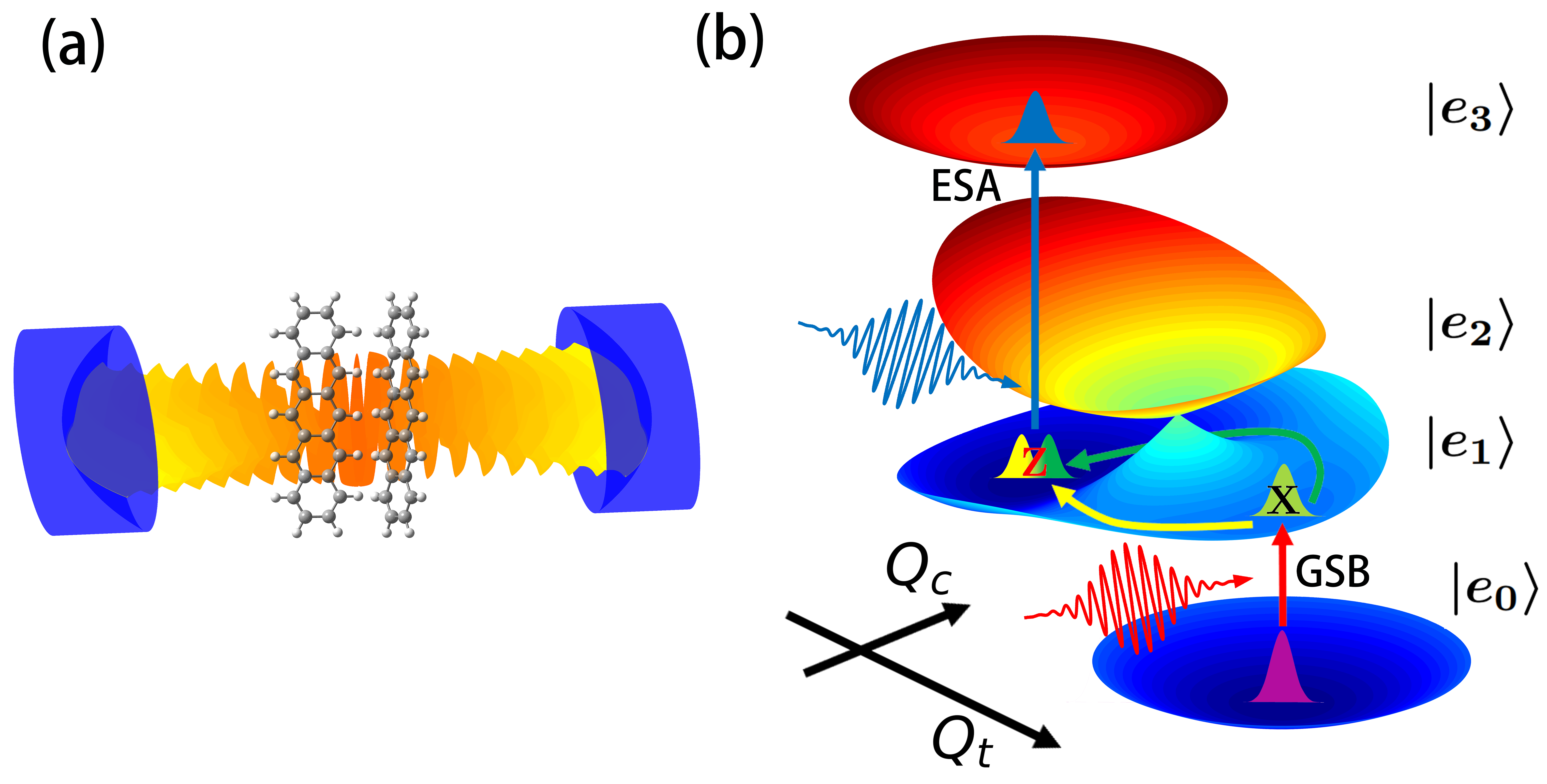}
\caption{\label{fig:Fig1}(a) Schematic representation of a molecular dimer coupled to an infrared cavity. (b) Conceptual energy landscape illustrating vibronic transitions and wave packet evolution. Here, $|e_{m^{\prime}}\rangle$ are four adiabatic states in $Q_t$-$Q_c$ space, X and Z are two key positions on surface $|e_{1}\rangle$. GSB and ESA denote the diagonal and off-diagonal signals of 2DES, respectively.}
\end{center}
\end{figure}

\section*{Results and Discussion} 

We consider a pentacene dimer for the study of the nonadiabatic dynamics and the associated GP near the CI (as illustrated in Fig.~\ref{fig:Fig1} (a)). It has been demonstrated that the primary singlet fission in pentacene undergoes an ultrafast deactivation of wave packet through the CI \cite{Duan Sci Adv 2020, Musser Nature Comm 2019, Kukura Nature Phys 2015}. To investigate the nonadiabatic dynamics near the CI, we consider four electronic states ($m$) of the pentacene dimer, i.e., the ground state S$_0$, the lowest singlet S$_1$, the correlated triplet-pair states $^{1}$TT and $^{2}$TT, and two vibrational modes, i.e., a tunning mode and a coupling mode. The tunning (coupling) mode, with frequency $\Omega_t$ ($\Omega_c$) and coordinate $Q_t$ ($Q_c$), modulates the energy $E_m$ of state $|m\rangle$ (the coupling between S$_1$ and $^{1}$TT) with a strength of $\kappa_m$ ($\lambda$). To study the cavity-modulated dynamics, an infrared optical cavity with coordinate $Q_a$ and frequency $\Omega_a$ (resonant with the coupling mode) is considered to interact with the coupling mode \cite{YangZhao JCP 2025, JPCL 9 1951 2018} via $\eta Q_cQ_a$. So the Hamiltonian for the system ($\hbar=1$) is $H_s = \sum_m (E_m + \kappa_m Q_t) |m\rangle\langle m| + \frac{1}{2} \sum_n \Omega_n (P^2_n + Q^2_n) + \lambda Q_c (|S_1\rangle \langle ^{1}TT| + H.c.) + \eta Q_c Q_a$, with $n=t, c, a$. The hybrid system is embedded in a bath composed of a set of harmonic oscillators. Considering the environmental effects, the quantum dynamics within the system is obtained using the time non-local (TNL) equations \cite{1999Non}. Detailed Hamiltonian and parameters are described in the Supporting information (SI).
 
\subsection{Wave packet dynamics of the molecular dimer}

Before investigating the wave packet propagation along the PESs, we first examine the landscape of the adiabatic PESs over the $Q_t$-$Q_c$ space \cite{Duan JCP 2017, Lipeng 2019} by diagonalizing $\langle Q_t|\langle Q_c|H_s|Q_c\rangle |Q_t\rangle$ with the four resulted adiabatic states $|e_{m^{\prime}}\rangle$ ($m^{\prime} = 0, 1, 2, 3$) schematically depicted in Fig.\ \ref{fig:Fig1}(b). A CI between the PESs of $|e_{1}\rangle$ and $|e_{2}\rangle$ can be clearly identified and plays a critical role in the subsequent dynamics. The nonadiabatic dynamics is initiated from a state located at the ``X'' point in Fig.\ \ref{fig:Fig1}(b), which is generated by the photo-excitation from the minimum of $|e_0\rangle$. Using the wave packet dynamics from the TNL equations, we obtain the time-resolved reduced density matrix and analyze the population in the reduced nuclear coordinate representation to elucidate the quantum dynamics across the PESs.

\begin{figure}[t!]
\begin{center}
\includegraphics[width=8.0cm]{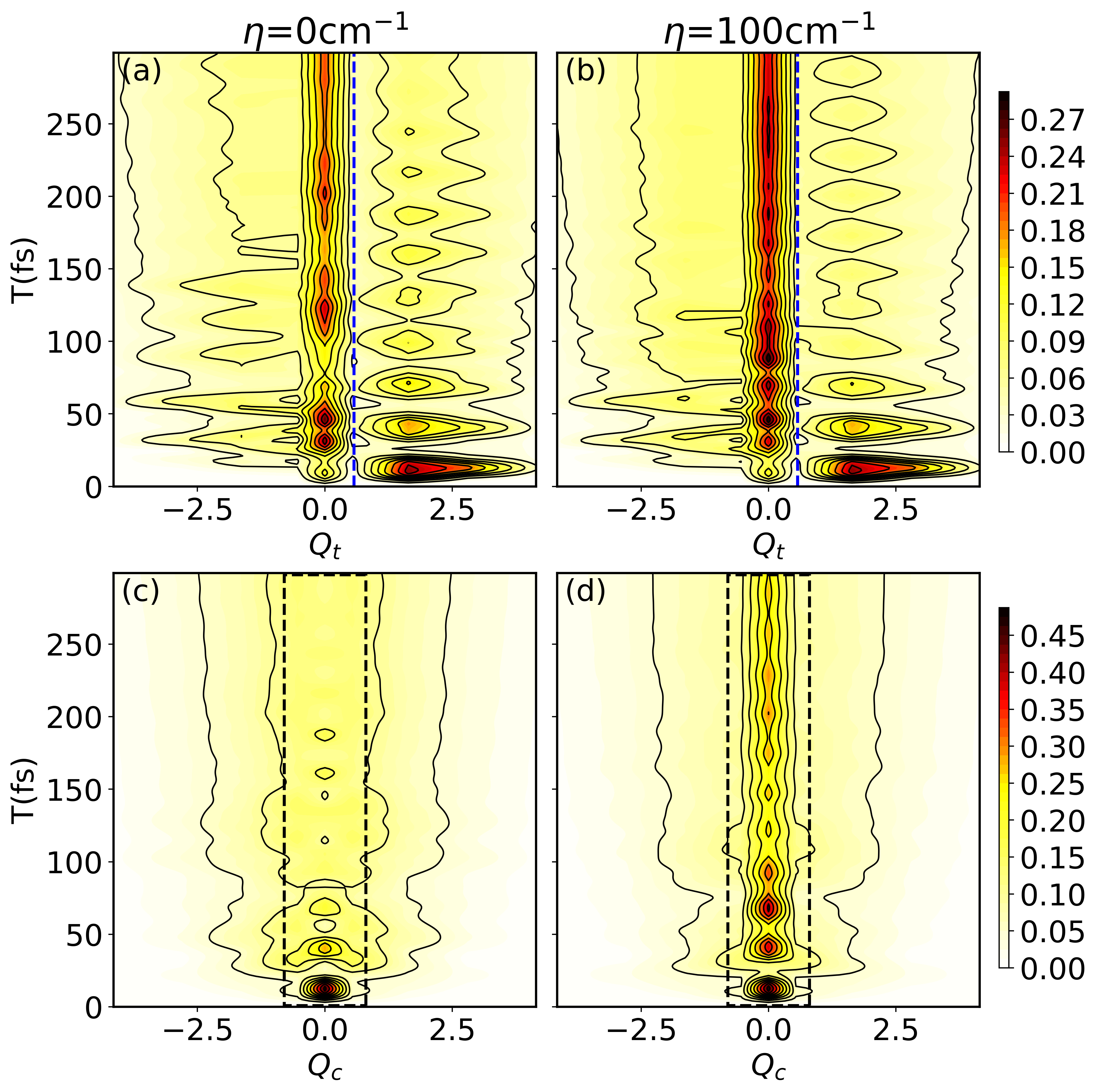}
\caption{\label{fig:Fig2}Sketch of the locations of three energy eigenstates on the energy landscape of $|e_1\rangle$ (a), with the wave function probability distributions shown in (b) for $\Psi^{E}_2$ and in (c) for $\Psi^{E}_3$ with $\eta$ = 50 cm$^{-1}$. Population dynamics on three eigenstates for $\eta$ = 0 (d) and $\eta$ = 50 cm$^{-1}$ (e) with the state energies (in cm$^{-1}$) presented in legends. Wave packet dynamics on $|e_1\rangle$ along $Q_c$ for $\eta$ = 0 (f) and $\eta$ = 50 cm$^{-1}$ (g). Integration is only performed over $Q_t<0$ (around  ``Z'') to exclude the portion of the initial state and highlight the cavity-induced behavior.
 }
\end{center}
\end{figure}

We also analyze the energy eigenstates of $H_s$ (obtained via $H_s|\Psi^E_m\rangle = \epsilon_m|\Psi^E_m\rangle$) and the population evolution on these states during wave packet dynamics, which can provide comprehensive insights for the 2DES signals. We pick three key eigenstates in the dynamics , i.e, $\Psi^{E}_1$, $\Psi^{E}_2$ and $\Psi^{E}_3$ as illustrated in Fig.\ \ref{fig:Fig2}(a). Their probabilities $|\Psi^{E}_i|^2$ are mainly localized at ``X'' (for $i=1$) and ``Z'' (for $i=2, 3$) of the $|e_1\rangle$ PES. The overall shape of $|\Psi^{E}_i|^2$ is conserved for varying coupling strength $\eta$ (see Fig. S3 in the SI). The initial state is dominated by $\Psi^E_1$, as both located at ``X''. Starting from this point, the wave packet evolves toward the CI, traverses it, and subsequently reaches the ``Z'' point. With the probability distribution associated with $|e_1\rangle$ shown in Fig.\ S3 (b) (for $\eta=0$) and Fig.\ \ref{fig:Fig2}(b) (for $\eta=50$ cm$^{-1}$), the second lowest eigenstate $\Psi^E_2$ exhibits a nodal line along $Q_c=0$. In contrast, the lowest eigenstate $\Psi^E_3$ is symmetrically distributed with respect to $Q_c$ (see Fig.\ S3 (c) for $\eta=0$ and Fig.\ \ref{fig:Fig2}(c) for $\eta=50$ cm$^{-1}$).

The wave packet starting from ``X'' could travel through either sides around the CI and finally reach ``Z''. In the absence of the cavity (i.e., $\eta$ = 0), the wave packets arriving at ``Z'' from different paths have a phase mismatch of $\pi$, which is the GP accumulated along the evolution. Thus, $\Psi^{E}_2$, instead of the lowest $\Psi^{E}_3$, gets populated in this case during the wave packet propagating from ``X'' to ``Z'', as shown in Fig.\ \ref{fig:Fig2}(d).

To visually monitor the wave packet propagation, the population on $|e_1\rangle$ along $Q_c$ with $\eta=0$ is presented in Fig.\ \ref{fig:Fig2} (f), which is obtained by integrating over the $Q_t<0$ region to eliminate the initial state contribution and highlight the dynamics at ``Z'' (details given in Sec. \Romannum{3} of SI). Owing to the GP, destructive interference reduces the wave packet amplitudes along $Q_c = 0$ for $\eta=0$, as presented in Fig.\ \ref{fig:Fig2} (f). This is the general feature of the pure relaxation dynamics around a CI and the wave packet propagation pathway is sketched in Fig.\ S4 (a) and denoted as Scheme $P_a$.

\subsection{Wave packet dynamics modulated by the cavity}

Comparing with the dynamics in the molecular dimer, we now highlight the cavity modulation on the wave packet dynamics by turning on the coupling strength $\eta$ between the cavity and the coupling mode. Although the distributions of the eigenstates $\Psi^{E}_i$ show weak dependence on $\eta$ (Fig.\ S3), the population dynamics on these states is modulated by the cavity, especially the dynamics on $\Psi^{E}_3$, as can be detected from the red lines in Fig.\ \ref{fig:Fig2} (d and e). It indicates that the cavity is actively involved in the wave packet propagation from ``X'' to ``Z''. The cavity effects can be understood by splitting the cavity-vibration coupling operator ($\eta \hat{Q}_c\hat{Q}_a$) into two parts $\hat{T}_1$ and $\hat{T}_2$, which acts on $\Psi(Q_c)$ and $\Psi(-Q_c)$ as $\hat{T}_1\Psi(+Q_c)=\eta \hat{Q}_c\hat{Q}_a\Psi(+Q_c)=\eta |Q_c|\Psi(+Q_c)$ and $\hat{T}_2\Psi(-Q_c)=\eta \hat{Q}_c\hat{Q}_a\Psi(-Q_c)= - \eta |Q_c| \Psi(-Q_c)$, respectively. Here, the effects of $Q_a$ are not considered in $T_1$ and $T_2$ for simplicity. Detailed descriptions are demonstrated in Fig. S4 and Eq. (S12). With a non-zero $\eta$, these two operations lead to a phase factor $\pi$, which cancels with the phase factor $\pi$ caused by the GP. In this scenario, the wave packets arriving at ``Z'' are in step and populate the lowest state $\Psi^{E}_3$. Assisted by the cavity, the wave packet cannot only propagate via this routine finally to $\Psi^{E}_3$ (red line in Fig.\ \ref{fig:Fig2} (e)), also can evolve along the aforementioned relaxation pathways  in Scheme $P_a$ targeting at $\Psi^{E}_2$ (green line in Fig.\ \ref{fig:Fig2} (e)). This hybrid scheme modulated by the cavity is denoted as Scheme $P_b$ (see Fig.\ S4 (b)). The two kinds of pathways from ``X'' to ``Z'' in Scheme $P_b$ are competitive. Increasing the cavity-vibration coupling $\eta$ facilitates the dynamics occurring through the cavity-assisted routine, as illustrated by the higher population on $\Psi^{E}_3$ in Fig. S5 for stronger $\eta$.

Such cavity induced effects are further highlighted in the wave packet dynamics around ``Z'' as shown in  Fig.\ \ref{fig:Fig2} (f) and Fig. S5. Compared with the vanishing amplitudes along $Q_c=0$ in the absence of the cavity, detectable portion of the wave packet resides in $Q_c=0$ with a finite $\eta$, mainly arriving at ``Z'' from the cavity-assisted pathway. The enhancement of wave packet amplitudes at $Q_{c} = 0$ indicates a disruption of the phase factor $\pi$ induced by the GP, suggesting that the cavity-vibration interaction perturbs the stability of the GP interference. A stronger coupling $\eta$ allows the cavity progressively diminishes the GP influence on the wave packet interference. Such effect is pronounced as time evolves, which can be seen from the long time dynamics presented in Fig. S5. Overall, these findings illustrate that the destructive interference of the wave packet at ``Z'' in Fig.\ \ref{fig:Fig1}(b), a hallmark of GP, can be effectively modulated by coupling the CI system to an optical cavity resonant with the coupling mode.

\subsection{Detecting GP by 2DES}

\begin{figure}[t!]
\begin{center}
\includegraphics[width=9.0cm]{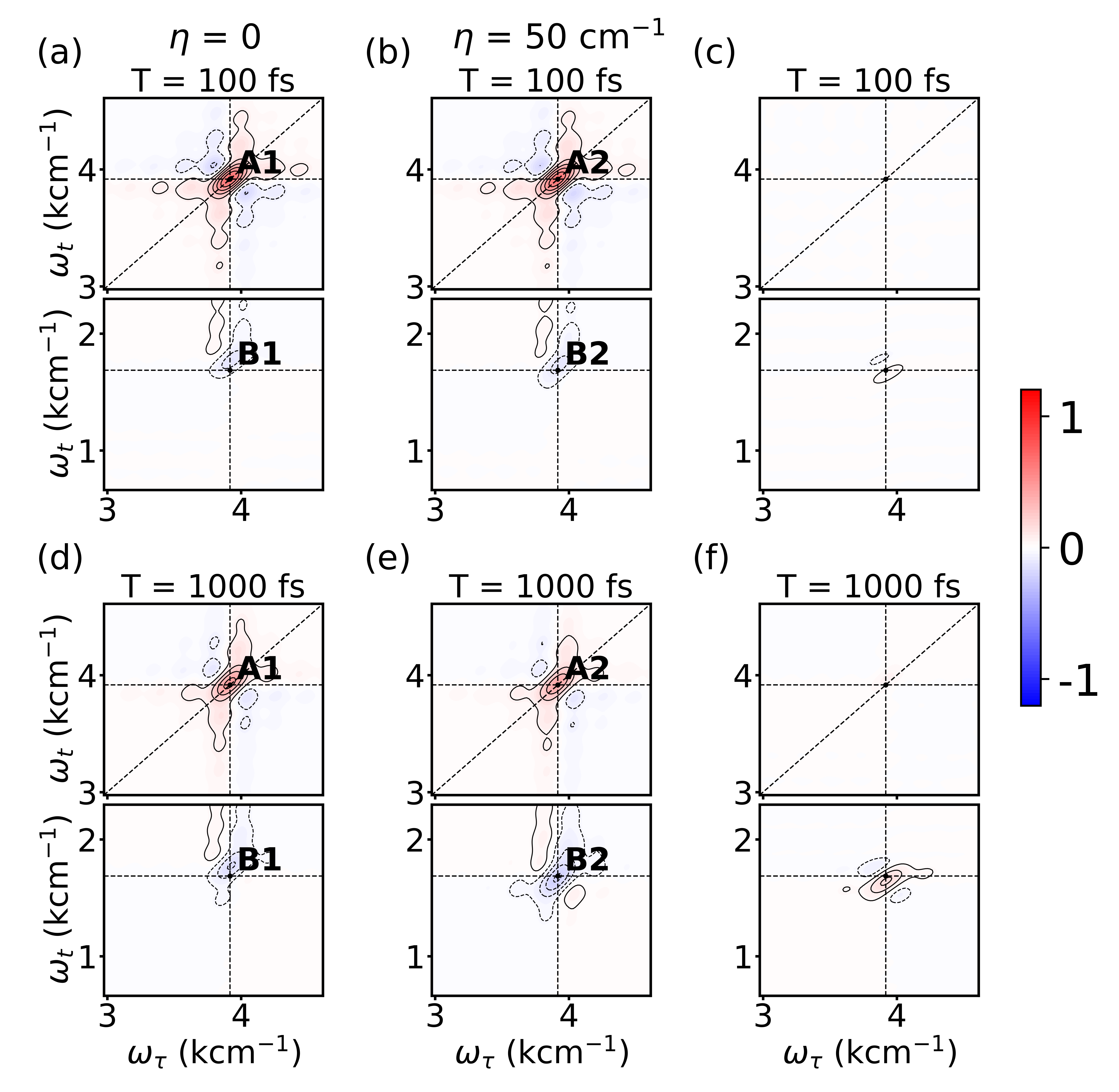}
\caption{\label{fig:Fig3}Calculated 2D electronic spectra with $\eta$ = 0 (a, d) and $\eta$ = 50 cm$^{-1}$(b, e). Their differential signals (subtracting the spectra of $\eta$ = 0 from that of $\eta$ = 50 cm$^{-1}$) are shown in (c, f). Each sub-figure contains two peaks A and B with A1 (A2) and B1 (B2) being the main and cross peaks for $\eta$ = 0 ($\eta$ = 50 cm$^{-1}$), respectively. Spectra at waiting times $T$ = 100 fs (a, b and c) and 1000 fs (d, e and f) are presented.
}
\end{center}
\end{figure}

With the knowledge of aforementioned wave packet dynamics around the CI, we propose a protocol here to precisely detect the quantum dynamics in such process using 2DES. Initializing the wave packet in the lowest eigenstate $|e_0\rangle$, we calculate the 2DES with various cavity-vibration coupling $\eta$ (details outlined in SI), and present the spectra for $\eta$ = 0 and 50 cm$^{-1}$ in the left (panels a and d) and middle (panels b and e) columns of Fig.\ \ref{fig:Fig3}, respectively. A diagonal positive peak (A1 for $\eta = 0$ and A2 for $\eta$ = 50 cm$^{-1}$) and an off-diagonal negative peak (B1 for $\eta = 0$ and B2 for $\eta$ = 50 cm$^{-1}$) are shown in each panel. The diagonal peaks (A1 and A2) are mainly contributed by the stimulated emission (SE) and ground-state bleach (GSB), while the cross peaks (B1 and B2) originate from the ESA. The differences between the spectra with different $\eta$ are also shown in the right column (panels c and f) of Fig.\ \ref{fig:Fig3}, vividly illustrating the cavity-modulated dynamics and the resulting spectra. Although the diagonal peaks are barely affected by the cavity, the ESA peaks exhibit apparent cavity-induced features, which are getting pronounced as the waiting time $T$ increases. It indicates that the cavity modulation on the wave packet propagation from ``X'' to ``Z'' can be directly detected via the ESA signals in 2DES.

To substantiate the interpretation of these features, we diagonalize the vibronic Hamiltonian ($H_s$) for both $\eta$ = 0 and $\eta$ = 50 cm$^{-1}$, and schematically present the transitions between the adiabatic states in Fig.\ S6. Although the absolute state energies are dependent on $\eta$, the $|\Psi^E_1\rangle$ - $|\Psi^E_0\rangle$ gap at ``X'' is not affected by the cavity, producing diagonal peaks (A1 and A2) located at $\omega_t = \omega_{\tau}$ = 3.953 kcm$^{-1}$, which are insensitive to $\eta$. Upon the initial excitation, the wave packet localized at ``X'' attempts to travel to ``Z'' where the ESA occurs. As demonstrated in the above section, the cavity introduces additional pathways for the wave packet propagation and produces different final wave packet configurations ($|\Psi^E_2\rangle$ and $|\Psi^E_3\rangle$) at ``Z'' (Scheme $P_b$). In contrast, in the absence of the cavity, the wave packet terminates at $|\Psi^E_2\rangle$ only (Scheme $P_a$) due to the GP accumulated in the conventional pathways around a CI (Figs.\ \ref{fig:Fig2}, S4, S5, and S6). Such different wave packet configurations at ``Z'' directly alter both the amplitudes and positions of the ESA peaks (B1 and B2), as highlighted in the right column of Fig.\ \ref{fig:Fig3}.

When $\eta$ = 0, GP-induced destructive interference cancels the wave packet amplitudes at ``Z'', thereby suppressing the ESA signal, which is mainly contributed by $|\Psi^E_2\rangle$ in Scheme $P_a$. Increasing $\eta$ activates $|\Psi_3\rangle$ and suppress the destructive interference through the additional propagation pathways (Scheme $P_b$), and in turn enhances the ESA intensity. Moreover, the cavity-vibration coupling $\eta$ decreases the energy gaps between the initial and final states for ESA, introducing lower-frequency components to the ESA peak. Such characteristic also gets prominent at longer times, as the population on $|\Psi^E_3\rangle$ increases as time evolves. Therefore, compared with B1, the peak B2 moves downward along $\omega_t$ as shown in Fig.\ \ref{fig:Fig3} (f).

This behavior is reserved across intermediate values of $\eta$, demonstrating the progressive recovery of the ESA peak with increasing coupling strength (Fig.\ S8). Comprehensive transition assignments and the associated eigenvalue analysis are provided in Section VIII of the SI (Fig.\ S6), confirming that the ESA signals (B1 and B2) in Fig.\ \ref{fig:Fig2} serve as a direct spectroscopic signature of the GP-induced interference and could be manipulated via the coupling to a cavity.

\begin{figure}[t!]
\begin{center}
\includegraphics[width=6.0cm]{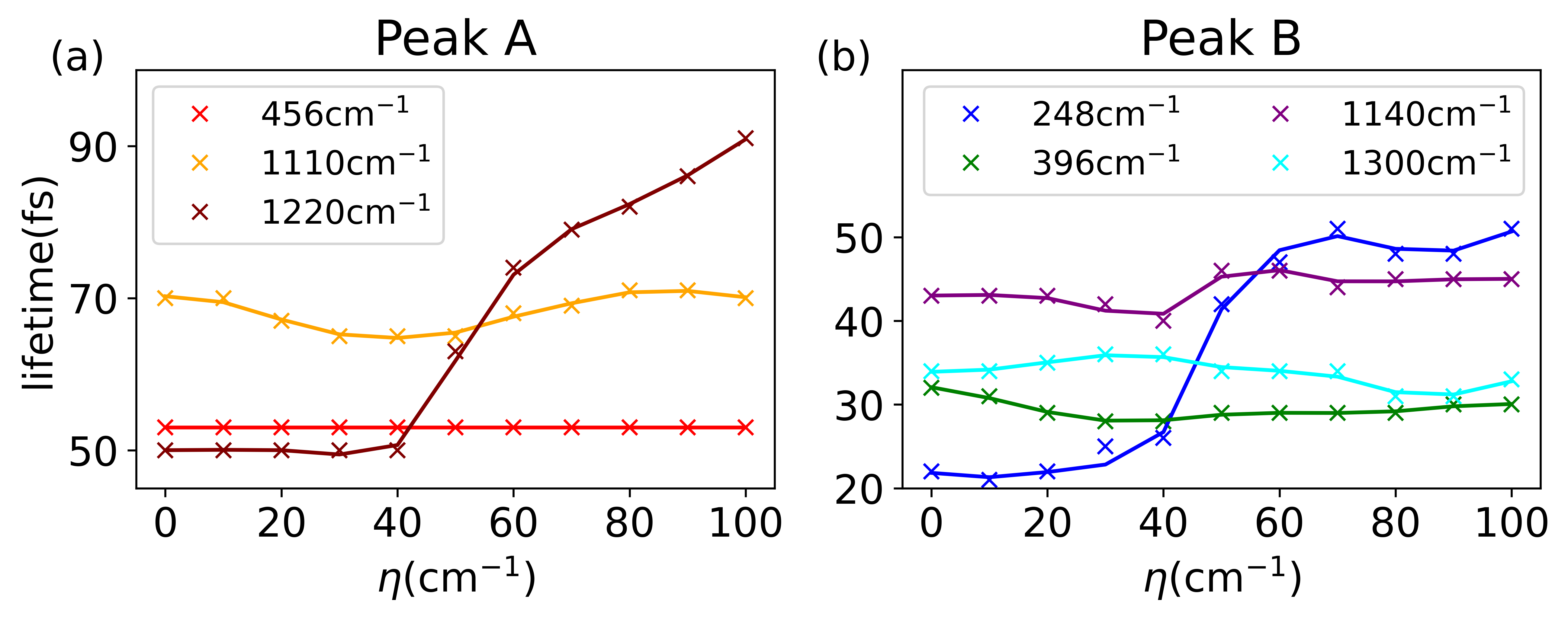}
\caption{\label{fig:Fig4}Intensities (dots) of peaks A1, A2, B1 and B2 in 2DES at different waiting times $T$ and corresponding exponential fitting $F(t) = a_f * e^{-t/\tau} + c_f$ (lines). Peaks A1 and A2 reflect GSB dynamics, while peaks B1 and B2 capture ESA behaviors. The corresponding fitting parameters $\tau$ are presented in legends.
}
\end{center}
\end{figure}

To gain deeper insights into population dynamics, we extend the 2DES simulations by increasing the waiting time ($T$). The time-resolved intensities of the four peaks (A1, A2, B1 and B2) in Fig.\ \ref{fig:Fig3} are presented as dots in Fig.\ \ref{fig:Fig4}. With an exponential fitting($F(t) = a_f * e^{-t/\tau} + c_f$, where $a_f$ and $c_f$ are constants), we obtain the lifetime $\tau$ numerically. Fitting of A1 and A2, associated with GSB, reveals that the lifetime slightly increases as $\eta$ increases from 0 to 50 cm$^{-1}$. In contrast, the lifetime from the fitting of peak B2 (156 fs) is almost double the one for B1 (79 fs), further indicating that the ESA signals are remarkably modulated by the cavity. These results underscore the role of cavity-enhanced vibronic interactions in modulating wave packet dynamics. In particular, the increase of the lifetime associated with peak B2 provides direct evidence of suppressed wave packet cancellation by the cavity. These findings demonstrate that 2DES can resolve and isolate spectral signatures linked to the GP near a CI, which could be further modulated by varying the coupling strength $\eta$. Additional data for $\eta$ = 25 cm$^{-1}$ and more computational details are provided in the SI. Overall, these results establish 2DES as a powerful tool for detecting GP-induced interference effects in strongly coupled light-matter systems.

\section*{Discussions} 

Cavity-modulated dynamics at CI offers a compelling approach for studying quantum coherent control in molecular systems. By combining 2DES with optical cavities, our work provides new understanding of topological effects in molecular wave packet evolution, especially the GP tied to CIs. We show that cavity-induced modulation can isolate spectroscopic signatures of the GP by selectively influencing wave packet pathways. Conventional transient absorption spectroscopy alone cannot resolve the GP signature identified here. The normal relaxations from ``X'' to ``Z'' without GP lead to final local lowest energy state whose feature like $\Psi^E_3$. But owing to the phase factor induced by GP the relaxations finally reach the local second lowest energy state whose feature like $\Psi^E_2$. When the CI system is coupled to an optical cavity it offers a path in another degree of freedom leading to $\Psi^E_3$. The local lowest energy state disturb the destructive pattern in wave packet dynamics, and the rate of this evolution controlled by cavity coupling contributes to a unique signature of ESA signals. 2DES disentangles excitation and detection frequencies, isolating cross-peaks and clearly revealing its component. Such frequency-resolved separation is thus crucial for exposing the interference arising from the GP.

The cavity’s spectral tunability, from visible to terahertz, enables precise manipulation of quantum dynamics. When coupled with the temporal resolution of 2DES, this setup allows detailed tracking of excited-state evolution. The strong nonadiabatic coupling near CI amplifies these quantum features, enhancing sensitivity to external perturbations. This creates opportunities not just for probing but actively steering molecular reactions. Such control is particularly valuable for processes like photoisomerization and singlet fission, which are central to biological vision, solar energy harvesting, and molecular optoelectronics. In realistic molecular systems such as crystalline pentacene, numerous vibrational modes couple to electronic states, disrupting the ideal symmetry of the potential-energy surfaces and obscuring GP interference. Hence, selecting molecules with simpler excited-state topology and well-defined tuning and coupling coordinates is essential.
 
Beyond molecular systems, this approach has strong parallels to emerging themes in quantum materials and topological photonics\cite{Rubio PNAS 2019, Appl Phys Rev 9 011312 (2022), Rubio Nature 2022}. Recent studies have shown that optical cavities can influence phase transitions, electronic structures, and correlated electron dynamics in solid-state systems\cite{Rubio Nature Mater 2020, PT Nature 2020}. Similarly, structured light and spatial light modulators are increasingly employed to control exotic states of matter, such as polariton condensates and Floquet topological insulators\cite{Rechtsman Science 2020}. Our work situates itself at the intersection of these fields, suggesting that cavity-modulated molecular systems can serve as analog platforms for studying broader quantum phenomena under controlled conditions. 

Looking forward, this research opens several compelling avenues. Experimentally, realizing cavity-modulated spectroscopy in the ultrafast regime will require precise cavity designs and tailored light-matter interaction schemes, but the technological progress in nanophotonics and cavity-QED makes this feasible. Theoretically, the inclusion of more complex vibrational modes, solvent environments, and entangled electronic states could further enrich the model and bring it closer to real molecular systems. Future efforts could also explore quantum entanglement generation, non-Hermitian dynamics near CIs, and quantum thermodynamic implications of topological phases in molecular reactions. In summay, our study provides a fundamental step toward unifying concepts of GP, nonadiabatic dynamics, and quantum control through the lens of cavity-modulated spectroscopy. This approach offers a versatile and deeply insightful platform for advancing both molecular physics and the broader field of quantum science.

\section*{Conclusions} 

Combining quantum modeling and simulations with 2DES, we propose a cavity-assisted strategy to detect the wave packet dynamics governed by the GP in the vicinity of a CI. We first investigate the nuclear wave packet propagation and population dynamics under varying cavity-vibration coupling, allowing us to distinguish the relaxation-only (Scheme $P_a$) and cavity-modulated-relaxation (Scheme $P_b$) pathways associated with the CI-mediated dynamics. The calculated 2DES  and comprehensive analysis provide a direct connection between the wave packet configuration near the CI and the ESA peak in 2DES. The spectra exhibit clear and robust spectroscopic signatures for the GP-induced destructive interference and cavity-induced modulations on the interference, demonstrating that the proposed cavity-coupled molecular system provides a reliable platform for detecting topological effects in nonadiabatic dynamics. 
Our study establish a coherent framework for probing GP-induced interference within CI-mediated relaxation pathways using multidimensional spectroscopy. Importantly, this work provides not only a theoretical protocol but also a potential experimental strategy for identifying GP signatures and thereby confirming the existence of CI-driven dynamics in molecular systems. More broadly, this approach opens new opportunities for investigating topological wave packet effects, coherent light-matter interactions, and cavity-controlled quantum dynamics in molecular and condensed-phase systems.

Acknowledgements
This work was supported by National Key Research and Development Program of China (Grant No.\ 2024YFA1409800),  NSFC Grant with No.\ 12274247 and 12504279, Zhejiang Provincial Natural Science Foundation of China with No.\ LQN26A040009, Yongjiang talents program with No. 2022A-094-G and 2023A-158-G, Ningbo International Science and Technology Cooperation with No.\ 2023H009, `Lixue+’Innovation Leading Project and the foundation of national excellent young scientist.

Author contributions: 
H. -G. D. conceived the research project. Y.C.Y. performed the calculations. H.-G.D. F. Z. and A.J. wrote the first draft of manuscript. All authors contributed towards the refinement of the manuscript. H.-G. D. supervised this project.   



\begin{thebibliography}{69}

\bibitem{Domcke book1} W. Domcke, D. R. Yarkony, H. K{\"o}ppel, Conical Intersections: Theory, Computation and Experiment; World Scientific: Singapore, 2011. 

\bibitem{Domcke book3} W. Domcke, D. R. Yarkony, Role of Conical Intersections in Molecular Spectroscopy and Photo-induced Chemical Dynamics. Annu. Rev. Phys. Chem. \textbf{63}, 325 (2012).  

\bibitem{Mukamle Chem Rev} M. Kowalewski, B. P Fingerhut, K. E Dorfman, K. Bennett, S. Mukamel, Simulating Coherent Multidimensional Spectroscopy of Nonadiabatic Molecular Processes: From the Infrared to the X-ray Regime. Chem. Rev. \textbf{117}, 12165 (2017). 

\bibitem{BO approx}  M. Born and R. Oppenheimer. Zur Quantentheorie der Molekeln. In: Annalen der Physik 389.20 pp. 457-484, (1927). 

\bibitem{NatPhys CI} Y. Chang, T. Balciunas, Z. Yin, M. Sapunar, B. N. C. Tenorio, A. C. Paul, and S. Tsuru. Electronic dynamics created at conical intersections and its dephasing in aqueous solution. Nature Physics \textbf{21}, 137-145 (2025). 

\bibitem{Massimo ChemSocRev} Y. Boeije, and M. Olivucci. From a one-mode to a multi-mode understanding of conical intersection mediated ultrafast organic photochemical reactions. Chem. Soc. Rev. \textbf{52}, 2643-2687 (2023).

\bibitem{PNAS Nucleobases} M. Barbatti, A. J. A. Aquino, J. J. Szymczak, D. Nachtigallová, P. Hobza, and H. Lischka. Relaxation mechanisms of UV-photoexcited DNA and RNA nucleobases. Proc. Natl. Acad. Sci. (USA) \textbf{107}, 21453-21458 (2010). 

\bibitem {ChemSci CI} A. Jha, {\em et al.}, Origin of poor doping efficiency in solution processed organic semiconductors. Chemical Science \textbf{9}, 4468-4476 (2018). 

\bibitem{Kukura Nature Phys 2015} A. J. Musser, M. Liebel, C. Schnedermann, T. Wende, T. B. Kehoe, A. Rao \& P. Kukura, Evidence for conical intersection dynamics mediating ultrafast singlet exciton fission. Nature Phys. \textbf{11}, 352 (2015). 

\bibitem{Duan Sci Adv 2020} H. -G. Duan {\em et al.} Intermolecular vibrations mediate ultrafast singlet fission. Science Adv. \textbf{6}, abb0052 (2020).  

\bibitem{Science Mathies 2005} P. Kukura, D. W. McCamant, S. Yoon, D. B. Wandschneider, and R. A. Mathies, Structural Observation of the Primary Isomerization in Vision with Femtosecond-Stimulated Raman. Science \textbf{310}, 1006 (2005). 

\bibitem{Science Tahara 2008} S. Takeuchi, S. Ruhman, T. Tsuneda, M. Chiba, T. Taketsugu, and T. Tahara, Spectroscopic Tracking of Structural Evolution in Ultrafast Stilbene Photoisomerization. Science \textbf{322}, 1073 (2008). 

\bibitem{Nature Chem Miller 2015} P. J. M. Johnson, A. Halpin, T. Morizumi, V. I. Prokhorenko, O. P. Ernst \& R. J. D. Miller, Local vibrational coherences drive the primary photochemistry of vision. Nature Chem. \textbf{7}, 980 (2015). 

\bibitem{JPCB Philip 2017} P. J. M. Johnson, M. H. Farag, A. Halpin, T. Morizumi, V. I. Prokhorenko, J. Knoester, T. L. C. Jansen, O. P. Ernst, R. J. D. Miller. The Primary Photochemistry of Vision Occurs at the Molecular Speed Limit. J. Phys. Chem. B \textbf{121}, 4040 (2017). 

\bibitem{Musser Nature Comm 2019} C. Schnedermann, A. M. Alvertis, T. Wende, S. Lukman, J. Feng, F. A. Y. N. Schröder, D. H. P. Turban, J. Wu, N. D. M. Hine, N. C. Greenham, A. W. Chin, A. Rao, P. Kukura \& A. J. Musser. A molecular movie of ultrafast singlet fission. Nature Comm. \textbf{10}, 4207 (2019). 

\bibitem{Matsumoto Nature Chem 2017} K. Miyata, Y. Kurashige, K. Watanabe, T. Sugimoto, S. Takahashi, S. Tanaka, J. Takeya, T. Yanai \& Y. Matsumoto, Coherent singlet fission activated by symmetry breaking. Nature Chem. \textbf{9}, 983 (2017). 

\bibitem {JPCL 12 2052 2021} B. Gu. {\em et al}. Optical-Cavity Manipulation of Conical Intersections and Singlet Fission in Pentacene Dimers. J. Phys. Chem. Lett. \textbf{12}, 2052 (2021). 

\bibitem {PCCP 23 16868 2021} M. H. Farag, {\em et al.} Polariton Induced Conical Intersection and Berry Phase. Phys. Chem. Chem. Phys. \textbf{23}, 16868 (2021). 

\bibitem {AVS Quant Sci 6 023501 2024} C. F{\'a}bri, {\em et al.} Classical and quantum light-induced non-adiabaticity in molecular systems. AVS Quant. Sci. \textbf{6}, 023501 (2024). 

\bibitem {JPCL 9 1951 2018} L. A. Martínez-Martínez,  {\em et al.} Polariton-Assisted Singlet Fission in Acene Aggregates. J. Phys. Chem. Lett. \textbf{9}, 1951 (2018). 

\bibitem{Domcke1} L. Chen, M. F. Gelin, V. Y. Chernyak, W. Domckeb and Y. Zhao, Dissipative dynamics at conical intersections: simulations with the hierarchy equations of motion method. Faraday Discussions, \textbf{194}, 61 (2006). 

\bibitem{Domcke2} S. Krempl, W. Domcke, M. Winterstetter, Real-time path-integral approach for general two-state multi-mode vibronic-coupling models. Chem. Phys. \textbf{206}, 63 (1996). 

\bibitem{Olivucci PNAS 2000} R. Gonz{\'a}lez-Luque, M. Garavelli, F. Bernardi, and M. Olivucci, Computational evidence in favor of a two-state, two-mode model of the retinal chromophore photoisomerization. Proc. Natl.Acad. Sci. (USA) \textbf{97}, 9379 (2000). 

\bibitem{Domcke3} L. Chen, M. F. Gelin, Y. Zhao, W. Domcke, Mapping of Wave Packet Dynamics at Conical Intersections by Time- and Frequency-Resolved Fluorescence Spectroscopy: A Computational Study. J. Phys. Chem. Lett. \textbf{10}, 5873 (2019). 

\bibitem{JCP 151 244102 2019} E. Mangaud, {\em et al.} Statistical distributions of the tuning and coupling collective modes at a conical intersection using the hierarchical equations of motion. J. Chem. Phys. \textbf{151}, 244102 (2019). 

\bibitem{JCP 151 014106 2019} A. J. Schile D. T. Limmer, Simulating conical intersection dynamics in the condensed phase with hybrid quantum master equations. J. Chem. Phys. \textbf{151}, 014106 (2019). 

\bibitem{Maxim JPB 2014} J. Krčmář, M. F. Gelin, D. Egorova and W. Domcke, Signatures of conical intersections in two-dimensional electronic spectra. J. Phys. B \textbf{47}, 124019 (2014). 

\bibitem{2DES1} E. Fresch {\em et al.} Two-dimensional electronic spectroscopy. Nat Rev Methods Primers \textbf{3}, 84 (2023).

\bibitem{2DES2} A. Jha {\em et al.} Quantum coherent dynamics in photosynthetic protein complexes. Chem. Soc. Rev. \textbf{55}, 1089-1130 (2026). 

\bibitem{Duan JPCL1} H. -G. Duan {\em et al.} Quantum Mechanical Wave Packet Dynamics at a Conical Intersection with Strong Vibrational Dissipation. J. Phys. Chem. Lett.  \textbf{7}, 382 (2016).

\bibitem{Duan JPCL2} H. -G. Duan {\em et al.} Impact of Vibrational Coherence on the Quantum Yield at a Conical Intersection. J. Phys. Chem. Lett.  \textbf{7}, 3491 (2016).

\bibitem{Egorova CP} M. Sala, D. Egorova. Two-dimensional photon-echo spectroscopy at a conical intersection: A two-mode pyrazine model with dissipation. Chem. Phys. \textbf{481}, 206 (2016). 

\bibitem{Mukamel1} D. Keefer, T. Schnappinger, R. de Vivie-Riedle,and S. Mukamel, Visualizing conical intersection passages via vibronic coherence maps generated by stimulated ultrafast X-ray Raman signals. Proc. Natl. Acad. Sci. (USA) \textbf{117}, 24069 (2020). 

\bibitem{Mukamel2} D. Cho and S. Mukamel, Stimulated X‑ray Raman Imaging of Conical Intersections. J. Phys. Chem. Lett. \textbf{11}, 33 (2020). 

\bibitem{Mukamel3} H. Yong, et al. Direct Monitoring of Conical Intersection Passage via Electronic Coherences in Twisted X-Ray Diffraction. Phys. Rev. Lett. \textbf{129}, 103001 (2022). 

\bibitem{Kowalewski1} D. Jadoun, M. Kowalewski, Time-Resolved Photoelectron Spectroscopy of Conical Intersections with Attosecond Pulse Trains. J. Phys. Chem. Lett. \textbf{12}, 8103 (2021). 

\bibitem{Mukamel4} M. Kowalewski, {\em et al.} Catching Conical Intersections in the Act: Monitoring Transient Electronic Coherences by Attosecond Stimulated X-Ray Raman Signals. Phys. Rev. Lett. \textbf{115}, 193003 (2015). 

\bibitem{Duan CP 2018} H. -G. Duan, D. -L. Qi, Z. -R. Sun, R. J. D. Miller, M. Thorwart. Signature of the geometric phase in the wave packet dynamics on hypersurfaces. Chem. Phys. \textbf{515}, 21 (2018). 

\bibitem {YangZhao JCP 2025} K. Sun, {\em et al.} Optical-cavity manipulation strategies of singlet fission systems mediated by conical intersections: Insights from fully quantum simulations. J. Chem. Phys. \textbf{162}, 130902 (2025). 

\bibitem {1999Non} C Meier, DJ Tannor. Non-Markovian evolution of the density operator in the presence of strong laser fields. J. Chem. Phys. \text{111} 3365 (1999).

\bibitem{Duan JCP 2017} D. Qi, {\em et al.} Tracking an electronic wave packet in the vicinity of a conical intersection. J. Chem. Phys. \textbf{147}, 074101 (2017). 

\bibitem{Lipeng 2019} L. Chen, M. F. Gelin, W. Domcke, Multimode quantum dynamics with multiple Davydov D2 trial states: Application to a 24-dimensional conical intersection model. J. Chem. Phys. \textbf{150}, 024101 (2019). 

\bibitem{Rubio PNAS 2019} C. Sch{\"a}fer, M. Ruggenthaler, H. Appel, and A. Rubio. Modification of excitation and charge transfer in cavity quantum-electrodynamical chemistry. Proc. Natl. Acad. Sci. (USA) \textbf{116}, 4883 (2019). 

\bibitem{Appl Phys Rev 9 011312 (2022)} F. Schlawin, D. M. Kennes, and M. A. Sentef. Cavity quantum materials. App. Phys. Rev. \textbf{9}, 011314 (2022). 

\bibitem{Rubio Nature 2022} J. Bloch, A. Cavalleri, V. Galitski, M. Hafezi \& A. Rubio, Strongly correlated electron-photon systems. Nature \textbf{606}, 41-48 (2022).

\bibitem{Rubio Nature Mater 2020} H. Hübener, U. De Giovannini, C. Schäfer, J. Andberger, M. Ruggenthaler, J. Faist \& A. Rubio, Engineering quantum materials with chiral optical cavities. Nature Mater. \textbf{20}, 438 (2020).

 \bibitem{PT Nature 2020} J. A. Muniz, D. Barberena, R. J. Lewis-Swan, D. J. Young, J. R. K. Cline, A. Maria Rey \& J. K. Thompson, Exploring dynamical phase transitions with cold atoms in an optical cavity. Nature \textbf{580}, 602-607 (2020).

\bibitem{Rechtsman Science 2020} S. Mukherjee \& M. C. Rechtsman, Observation of Floquet solitons in a topological band gap. Science \textbf{368}, 856-859 (2020).

\bibitem {1995principles} S. Mukamel. Principles of nonlinear optical spectroscopy, Oxford University Press, New York. (1995). 



\end{thebibliography}


\end{document}